\documentstyle[aps,eqsecnum]{revtex}
\title{Asymptotic expansion in the fission process}
\author{Asish K. Dhara, Sailajananda Bhattacharya and Kewal Krishan}
\address{Variable Energy Cyclotron Centre, 1/AF Bidhan Nagar, Calcutta
-700064,India}
\tighten
\begin{document}

\draft

\maketitle
\date{\today}
\begin{abstract}
Asymptotic  expansion  of  the  Fokker-Planck  equation  in  terms of the
strength of the fluctuation has  been  carried  out.  The  mean  and  the
variance of the total kinetic energies of the fission fragments have been
calculated and compared with the experimental values.

\end{abstract}

\section{introduction}
At  present  it  is  commonly  agreed  upon that the fission process is a
dissipative phenomena, where initial energy of the  collective  variables
get  dissipated  into  the  internal  degrees of freedom of nuclear fluid
giving rise  to  the  increase  in  internal  excitation  energy  causing
evaporation  of  a  large  number of precision neutrons, which can not be
explained by standard statistical model calculations. As  dissipation  is
referred  to  the  interaction  of  the  system coordinate with the large
number of degrees of freedom of the surrounding reservoir,  this  process
is   always   associated   with  the  fluctuation  of  relevant  physical
observable.  This  mesoscopic  description   is   inevitable   once   the
fluctuations of the observables are amenable to experimental observation.

Keeping  this fact in mind, the fission process is picturised as follows;
the collective variable such as  elongation  axis  acts  as  a  'Brownian
particle'  interacting  stochastically  with  large  number  of  internal
nucleonic degrees of freedom constituting the surrounding 'bath'. Several
workers solve either  the  Langevin  equation  \cite{abe1,abe2,frob},  or
multidimensional   Fokker-Planck  equation  \cite{fp1,fp2},  which  is  a
differential version of Langevin equation, in order  to  study  the  time
evolution  of  probability  distribution function and have calculated the
mean as well as the variance of the total kinetic energies of the fission
fragments.

The experimental data show that the root mean square deviation of kinetic
energy  is  always very small compared to the mean kinetic energy ($\sim$
0.1). The question naturally arises whether one can utilise  this  simple
fact  in  the theoretical scheme instead of solving the Langevin equation
or corresponding Fokker-Planck equation in detail.  In  this  spirit,  we
assume  that  the  full  solution of the Fokker-Planck equation admits an
asymptotic expansion in terms of strength of the fluctuation. Admittedly,
in the zeroth order of fluctuation one should obtain macroscopic behavior
of the dynamical evolution. Obviously, when the  fluctuation  is  ignored
one  obtains  the  deterministic picture and the machinery to handle such
situation is to solve the Euler-Lagrange equation  of  motion.  This  was
done  in  our  previous works \cite{dha2,dha3,dha}, where dissipation was
generated through nonconservative Rayleigh function and the total kinetic
energies, the fission yields and neutron multiplicities were  calculated.
However, in the above scheme, there had been no scope to study the effect
of  fluctuations  originating  from the stochastic dynamics of fission on
fission observables. Here, we report a modified scheme where fluctuations
have been included in the fission dynamics by making asymptotic expansion
of  the  probability  distribution  function  in  terms  of  intensity or
strength of the fluctuations as argued before. Thus, the picture we adopt
here, is as follows: Due to smallness of the  relative  fluctuation,  the
process  is  grossly  described  in  terms  of  macroscopic equation. The
stochastic description is introduced by studying  the  evolution  of  the
probability distribution of narrow width over its deterministic values.

In  what  follows, in Sec.II we describe briefly the procedure and derive
the equations of  the  moments.  The  results  of  the  calculations  are
discussed in Sec.III. Finally, concluding remarks are given in Sec.IV.

\section{Asymptotic expansion of the Fokker-Planck equation}
\subsection{The method}

As argued before, the mesoscopic description of the fission process begins
with a set of Langevin equations:

\begin{mathletters}
\begin{eqnarray}
\label{eq.1}
\dot{X_i} &=& h_i(\{X\},\{Y\}) +\eta_i(t)   \\
\dot{Y_i} &=& H_i(\{X\},\{Y\})  \hspace{18pt} ; i=1,2,3,...,N
\end{eqnarray}
\end{mathletters}

where  $h_i$  and  $H_i$ are given functions of the stochastic collective
variables $ X_1,X_2,...,X_N $ and $  Y_1,Y_2,...,Y_N  $  in  the  fission
process and $ \eta_i(t)$ refers to the driving noise term associated with
the  interaction  of the {\it i}th collective variable with the reservoir
constituting nucleonic degrees of freedom. For simplicity, we assume  the
noise  to  be a gaussian white with zero mean and decoupled for different
degrees of freedom with auto-correlation functions given by

\begin{equation}
\label{eq.2}
<\eta_i(t)> = 0,\hspace{8pt} <\eta_i(t)\eta_j(t')> = D_i(y_i)\delta(t-t')
\delta_{ij}  ,
\end{equation}

where  $D_i(y_i)$  is the diffusion coefficient associated with {\it i}th
variable, depending only on the sample space $y_i$ for the stochastic
 variable $Y_i$.

The Fokker-Planck equation corresponding to the Langevin equation (2.1) is

\begin{equation}
\label{eq.3}
\frac{\partial f(\{x\},\{y\},t)}{\partial t}  =  -\sum_i [\frac{\partial
(h_i f)} {\partial x_i} + \frac{\partial (H_i  f)}{\partial  y_i}  -(1/2)
D_i(y_i) \frac{\partial^2 f}{\partial x_i^2}]   .
\end{equation}
The  quantity  $f(\{x\},\{y\},t)$  is  the  probability  density function
depending on the variables $ x_1,x_2,...,x_N, y_1,y_2,...,y_N $ and  time
$t$ explicitly. If we are interested in finding the time evolution of the
conditional  probability  distribution  function  then  we  have to solve
Eq.(2.3)   with   initial    values    $x_i(0)=x_i^0$,    $y_i(0)=y_i^0$,
$\forall$$i$,  at  $t=0$.  That  is,  we have to solve Eq.(2.3) for those
realisations which are known to start from these specific points  in  the
whole sample space.

The  asymptotic  expansion  was  developed  by  van Kampen \cite{van} for
constant  diffusion  constant.  The  method  consists  of   writing   the
stochastic  variables as the sum of deterministic value and a fluctuating
part at each time $t$ with root of the diffusion constant as  a  strength
of  the  fluctuating part. In the present paper we generalise this method
for the  situations  where  the  diffusion  coefficients  depend  on  the
stochastic  variables  explicitly. Such a situation is encountered in the
case  of  fission  process,  where  the  friction   coefficient   depends
explicitly  on  the  collective  variable or shape of the nucleus at each
instant of time. In this case, we further assume that, in the  asymptotic
expansion,  the  strengths  of  the  fluctuating  parts of the stochastic
variables depend only on the deterministic values of the  respective  $y$
variables:

\begin{mathletters}
\begin{eqnarray}
\label{eq.4}
x_i&=&\bar{x_i} + \surd D(\bar{y_i}) \zeta_i\\
y_i&=&\bar{y_i} + \surd D(\bar{y_i}) \xi_i
\end{eqnarray}
\end{mathletters}

The  quantities  $\{\zeta_i\},\{\xi_i\}$ refer to the fluctuations of the
stochastic variables $\{x_i\}$ and $\{y_i\}$ around  their  deterministic
values   $\{\bar{x_i}\},  \{\bar{y_i}\}$.  Next,  we  introduce  the  new
distribution   function   $Q$   depending   only   on    the    variables
$\{\zeta_i\}$,$\{\xi_i\}$  and  $t$. The normalisation condition suggests
that the $f$ and $Q$ will be related by

\begin{equation}
\label{eq.5}
f(\{x\},\{y\},t)  =  {\prod_{i=1}^N[D_i (\bar{y_i})]^{-1}} Q(\{\zeta\},
\{\xi\},t)
\end{equation}

Substituting  Eq.(2.4)  in  the Fokker-Planck equation(2.3),making Taylor
expansion of $h(\{x\},  \{y\}),  H(\{x\},  \{y\})$  around  $\{\bar{x}\},
\{\bar{y}\}$   and   collecting   coefficients   of   various   order  of
$D(\bar{y_i})$,we could generate a hierarchy of equations.As expected,the
first set would give rise to the equation of motion for $\{\bar{x}\}$ and
$\{\bar{y}\}$.

\begin{mathletters}
\begin{eqnarray}
\label{eq.6}
\dot{\bar{x_i}} &=& h_i(\{\bar{x}\},\{\bar{y}\})  \\
\dot{\bar{y_i}} &=& H_i(\{\bar{x}\},\{\bar{y}\}) \hspace{18pt};\forall{i}
\end{eqnarray}
\end{mathletters}

Eqs.(2.6)  are the Euler-Lagrange equation for deterministic motion.These
equations are to  be  solved  with  initial  conditions  $\{\bar{x}(0)\}=
\{x^0\},  \{\bar{y}(0)\}=  \{y^0\}$.  Next, we are going to calculate the
conditional  probability   distribution   $f(\{x\},\{y\},   t\mid\{x^0\},
\{y^0\},0)$ or $Q(\{\zeta\}, \{\xi\}, t\mid0,0,0)$.

Assuming  the variation of diffusion coefficient over the narrow width of
the distribution function at any instant of time to be $\bigcirc(D)$,  we
could replace the second Fokker-Planck coefficient $D(y)$ by $D(\bar{y})$
at  each  instant of time.This assumption makes the calculation extremely
simple.Collecting coefficients $\bigcirc(D^0)$, we get  back  quasilinear
Fokker-Planck equation for $Q$:

\begin{equation}
\label{eq.7}
\frac{\partial Q}{\partial t}+\sum_i(\frac{\dot{D}(\bar{y}_i)}
{D(\bar{y}_i)})Q
=-\sum_i[a_i\frac{\partial(\zeta_i Q)}{\partial\zeta_i}
        +b_i\frac{\partial(\xi_i Q)}{\partial\zeta_i}
        +c_i\frac{\partial(\xi_i Q)}{\partial\xi_i}
        +d_i\frac{\partial(\zeta_i Q)}{\partial\xi_i}
        -(1/2)\frac{\partial^2Q}{\partial\zeta_i^2}]
\end{equation}

where $a_i,b_i,c_i,d_i$ are given by
\begin{mathletters}
\begin{eqnarray}
\label{eq.8}
a_i&=&(\frac{\partial  h}{\partial\bar{x}_i})-  (\frac{\dot{D}(\bar{y}_i)
}{2D(\bar{y}_i)})\\
b_i&=&(\frac{\partial h}{\partial\bar{y}_i})\\
c_i&=&(\frac{\partial H}{\partial\bar{y}_i}) - (\frac{\dot{D}(\bar{y}_i)}
{2D(\bar{y}_i)})\\
d_i&=&(\frac{\partial H}{\partial\bar{x}_i})
\end{eqnarray}
\end{mathletters}

Eq.(2.7) suggests that
\begin{equation}
\label{eq.9}
Q(\{\zeta\},\{\xi\},t) = \prod_jQ_j(\zeta_j,\xi_j,t)
\end{equation}

where  the distribution function $Q_j$ for each $j$ satisfies the similar
equation written below without the subscript:

\begin{equation}
\label{eq.10}
\frac{\partial Q}{\partial t}+(\frac{\dot{D}(\bar{y})}
{D(\bar{y})})Q
=-[a\frac{\partial(\zeta Q)}{\partial\zeta}
        +b\frac{\partial(\xi Q)}{\partial\zeta}
        +c\frac{\partial(\xi Q)}{\partial\xi}
        +d\frac{\partial(\zeta Q)}{\partial\xi}
        -(1/2)\frac{\partial^2Q}{\partial\zeta^2}]
\end{equation}

subject to the initial condition

\begin{equation}
\label{eq.11}
Q(\zeta,\xi,t=0) = \delta(\zeta)\delta(\xi)
\end{equation}

The solution of Eq.(2.10) is given by

\begin{equation}
\label{eq.12}
Q(\zeta,\xi,t)  =  [\frac{1}{(2\pi)^2}]\int\int  exp-\{ik\zeta+il\xi  +
\frac{ [g(t)k^2+G(t)l^2 + 2c(t)kl]}{2D(t)}\} dk dl
\end{equation}

where   $g(t),G(t),c(t)$   satisfy   the   set  of  coupled  first  order
differential equations :

\begin{mathletters}
\begin{eqnarray}
\label{eq.13}
\frac{\dot{g}}{2}&  = &(\frac{\partial h}{\partial x})g + (\frac{\partial
h}{\partial    y})c+\frac{D}{2}\\
\frac{\dot{G}}{2}&=&(\frac{\partial
H}{\partial y})G+(\frac{\partial H}{\partial x})c\\
 \dot{c}  &=&(\frac{\partial  h}  {\partial  x})c  +  (\frac{\partial  h}
{\partial y})G+ (\frac{\partial H}{\partial  y})c  +  (\frac{\partial  H}
{\partial x})g
\end{eqnarray}
\end{mathletters}

with the initial conditions

\begin{equation}
\label{eq.14}
g(0)=G(0)=c(0)=0
\end{equation}

Once  $Q(\zeta,\xi,t)$  is  known,from  Eq.(2.9)  and  Eq.(2.5)  the full
conditional   probability   distribution   function   $f(\{x\},    \{y\},
t\mid\{x^0\},  \{y^0\},0)$  is  known. Integrating this function over all
variables except one,say $x_i$, one identifies $g_i(t)$ as  the  variance
of the stochastic variable $X_i$.

\begin{equation}
\label{eq.15}
<(X_i-<X_i>)^2> = g_i(t)
\end{equation}

We  note  that  the  homogeniety of Eq.(2.10) suggests that $<\zeta(t)> =
<\xi(t)>=0$, or the average of the variables $X$ and $Y$ at any time will
be  determined  by  their  deterministic  values  obtained   by   solving
Euler-Lagrange equation (2.6). Similarly one observes from Eq.(2.12),

\begin{mathletters}
\begin{eqnarray}
\label{eq.16}
<(X_i-<X_i>)><(Y_i-<Y_i>)>&=& c_i(t) \\
<(Y_i-<Y_i>)^2>&=& G_i(t)
\end{eqnarray}
\end{mathletters}

\subsection{Application to the fission process}

In  the  fission  process,we  choose  the  elongation  axis  and velocity
associated with it as the stochastic variables interacting with  a  large
number  of internal nucleonic degrees of freedom constituting a heat bath
at temperature $T$ determined by the excitation energy available  to  it.
We  further  assume  that  the  'collisional' time scale of the nucleonic
degrees of freedom is much shorter than the time scale of the macroscopic
evolution of the collective variable so that at each instant of time  the
heat bath is assumed to be in quasi-stationary equilibrium.

As  stated  before  the  Euler-Lagrange  equation(2.6)  was solved in our
earlier works \cite{dha}. To avoid repetition we  deliberately  omit  the
procedure and scheme to solve that equation. For the sake of completeness
we  merely write that equation and refer to our previous paper to clarify
the details.

Giving correspondence to the terminology used in this paper, we associate

\begin{equation}
\label{eq.17}
Y=r,X=\dot{r}
\end{equation}

Thus we have

\begin{mathletters}
\begin{eqnarray}
\label{eq.18}
H(x,y) &=& x =\dot{r}\\
h(r,\dot{r})&    =    &[\frac{L^2}{\mu    r^3}    -    \gamma\dot{r}    -
\frac{\partial(V_C+V_N)} {\partial r}]/\mu
\end{eqnarray}
\end{mathletters}

The  quantities  $V_C,V_N$  represent the Coulomb and nuclear interaction
potentials, respectively. The quantity $\gamma, \mu$  and  $L$  refer  to
friction  coefficient, reduced mass associated with the fissioning liquid
drop and the relative angular momentum, respectively\cite{dha}.

While  solving  for  variances  we  appeal  to Eq.(2.13). In that set the
diffusion coefficient $D$ is evaluated employing  Einstein's  fluctuation
dissipation theorem. The set thus reads

\begin{mathletters}
\begin{eqnarray}
\dot{g}(t)&=&2(\frac{\partial h}{\partial \dot{r}})g(t)+
2(\frac{\partial h}{\partial r})c(t)+
2\gamma(r)T(r)/ \mu^2\\
\dot{G}(t)&=& 2c(t)\\
\dot{c}(t)&=&(\frac{\partial h}{\partial \dot{r}})c(t)+
2(\frac{\partial h}{\partial r})G(t)+g(t)
\label{eq.19}
\end{eqnarray}
\end{mathletters}

with  the  initial  conditions(2.14).  The  initial conditions of $r$ and
$\dot{r}$ for solving Eq.(2.18) are\cite{dha}

\begin{equation}
\label{eq.20}
r(t=0)=r_{min},\dot{r}(t=0)=(\frac{E^*R_N}{2\mu})^{1/2}
\end{equation}

where  the  fissioning  nucleus  starts from the minimum of the potential
energy surface, the quantity $R_N$ is a random number  between  0  and  1
from uniform probability distribution and $E^*$ is the available energy.

Solving  Eq.(2.18)  and Eqs.(2.19) simultaneously with initial conditions
(2.14) and (2.20) we generate the  conditional  probability  distribution
function  $f(r,\dot{r},t\mid  r(t=0),  \dot{r}(t=0), 0)$. The probability
distribution function $f(r,\dot{r},t)$ could be obtained as

\begin{equation}
\label{eq.21}
f(r,\dot{r},t) = \int f(r,\dot{r},t\mid r(t=0),\dot{r}(t=0),0)
                      f(r(t=0),\dot{r}(t=0),0)
                      dr(t=0)d\dot{r}(t=0)
\end{equation}

where  $f(r(t=0),\dot{r}(t=0),  0)$  is  the  probability distribution of
position and velocity of the stochastic variables at the initial time. As
described by the initial condition(2.20), this can be represented as

\begin{equation}
\label{eq.22}
f(r(t=0), \dot{r}(t=0), 0) = \delta(r(t=0)-r_{min})\times f(\dot{r}(t=0))
\end{equation}

Here, we assumed that each fissioning nucleus in the ensemble starts from
a  fixed  initial  position  but  with  different  partioning  of initial
excitation  energy\cite{dha}.  Finally,  substitution  of  Eq.(2.22)   in
Eq.(2.21) would give

\begin{equation}
\label{eq.23}
f(r,\dot{r},t)        =        \sum_{R_N}f(r,\dot{r},t\mid       r_{min},
(\frac{E^*R_N}{2\mu})^{1/2},0)
\end{equation}

\section{Results and discussions}

The temporal evolutions of the variables $g(t)$, $c(t)$, $G(t)$ along the
fission  trajectory  have been computed by solving numerically the set of
Eqs.~\ref{eq.19}. The  results  are  plotted  in  Fig.~\ref{fig1}  for  a
representative  system  $^{16}$O + $^{124}$Sn. It is seen from the figure
that, initially, all of them increase steeply and then  their  magnitudes
become   nearly   constant   throughout   the  rest  of  the  trajectory.
Furthermore, the calculation shows that

\begin{equation}
\label{eq.3.1}
\frac{c^2(t)}{g(t)G(t)}\ll 1
\end{equation}

This  implies  that  the  correlation  of  position  and  velocity of the
elongation variable $(r)$ is much smaller compared  to  their  respective
variances.  This fact simplifies our calculation of variance of energy of
the fission fragments. The variance of  energy  and  average  of  kinetic
energy at scission point are given by

\begin{mathletters}
\begin{eqnarray}
\label{eq.3.2}
\sigma_E^2&  =  &(\mu\dot{r})^2g(t) + [\frac{\partial(V_C+V_N)} {\partial
r}]^2G(t)\\
<E(t)>&=&\mu g(t_{sc})/2+E_{det}
\end{eqnarray}
\end{mathletters}

where  $t_{sc}$  is  the  time  at  scission  point  and $E_{det}$ is the
deterministic value of total fragment kinetic energy (TKE) after scission
and $\sim$~ 100 - 200 MeV. In  deriving  the  above  approximate  results
Eqs.~(3.2),  we utilise our observation (3.1). It is further assumed that
the variation of the potential over the narrow width of  the  probability
distribution   is   small  so  that  the  average  of  the  potential  is
approximated as the value of the potential  at  the  mean  position.  The
variation  of  the  kinetic energy variance $\sigma_E^2$ as a function of
time  has  also  been  displayed  in  Fig.~\ref{fig1}.   The   value   of
$\sigma_E^2$  is  also seen to increase steeply at the beginning and then
it becomes nearly constant throughout the rest of the time. As  envisaged
earlier,  the result clearly shows that $\frac{\sigma_E}{<E>}\ll1$, which
demonstrates the validity of asymptotic expansion in deriving the  result
instead of solving the Fokker-Planck equation in detail.

The  theoretical predictions of $\sigma_E(th)$ for the fission of several
compound systems produced in the 158.8 MeV $^{18}$O and 288 MeV  $^{16}$O
reactions  on  various  targets  have  been  displayed in Fig.~\ref{fig2}
alongwith the respective experimental data \cite{hinde}  for  comparison.
The  experimental  data  $\sigma_E(exp)$  is  represented  by  the filled
circles and the solid curves are the results  of  the  present  calculati
ons. It is seen from Fig.~\ref{fig2} that when the projectile energy (and
vis-a-vis  the  excitation  energy  of the fused composite) is relatively
lower (lower half ), the calculated values are in fair agreement with the
data. However, the calculation underpredicts the  experimental  value  of
$\sigma_E$  for  the heaviest target considered ($^{238}$U in the present
case). With the increase in the projectile  energy  (and  the  excitation
energy of the composite) (upper half of Fig.~\ref{fig2}), the theoretical
predictions  are  found  to  underestimate the corresponding experimental
values and the discrepancy between the two increases with the increase in
mass number. We have also studied the fragment mass asymmetry  dependence
of  energy  variance,  $\sigma_E^2(A_1,  A_2)$  for  some  representative
systems and the results are displayed in Fig.~\ref{sig_asym}. It is  seen
from  Fig.~\ref{sig_asym}  that  the theoretical values of variances have
only a weak dependence on the fragment mass asymmetry.

In order to investgate into the discrepancy between the predicted and the
experimental  values  of  the  TKE  variance,  it  is  observed  that the
experimental values of $\sigma_E(exp)$ are usually obtained by  averaging
over the full mass yield spectrum. Therefore, $\sigma_E(exp)$ consists of
two  terms, {\it viz.}, ({\it i}) contributions arising due to stochastic
fluctuations  in  the  dynamics  of  fission  process  and   ({\it   ii})
contributions  from  the  variation  of  the mean kinetic energy with the
fragment  mass  asymmetry.  So,  $\sigma_E(exp)$  may   be   written   as
\cite{laza},

\begin{mathletters}
\begin{eqnarray}
\label{eq:sigy}
\sigma_E^2(exp) = \sigma_E^2 + \sigma_E^2(kin) \\
\sigma_E^2 =\sum_{A_1}  \sigma_E^2(A_1, A_2) \cdot Y(A_1) \\
\sigma_E^2(kin) =\sum_{A_1}[ \bar{E} - <E(A_1,A_2)> ]^2  \cdot Y(A_1)
\end{eqnarray}
\end{mathletters}

where  $\sigma_E^2(A_1,  A_2)$  and  $<E(A_1,A_2)>$ are the variances and
mean values of the total kinetic energy of  two  fission  fragments  with
mass  numbers  $A_1$  and  $A_2$  (compound nucleus mass $ A_{CN} = A_1 +
A_2)$,  $\bar{E}$ being the average of $<E(A_1,A_2)>$ over the normalised
fragment mass yield, $Y(A_1)$  with  $\sum_{A_1}  Y(A_1)  =  1$.  It  is,
therefore,  clear  that  for  a  proper  comparison  of  the  theoretical
predictions of kinetic energy variance  with  the  relevent  experimental
data,  the  theoretical  numbers  should  be averaged over the respective
fragment  mass  distribution.   Moreover,   the   calculated   value   of
$\sigma_E^2(th)$ should be compared with the extracted experimental value
of   $\sigma_E^2$   obtained   by   substracting  $\sigma_E^2(kin)$  from
$\sigma_E^2(exp)$.

 We  have  calculated  $\sigma_E^2(kin)$  for a few systems for which the
experimental fragment mass yield data are available \cite{hinde},  taking
$<E(A_1,A_2)>$ from Viola systematics \cite{viola}. The results are given
in Table~\ref{t1} alongwith the values of $\sigma_E^2$ extracted from the
experimental  data.  It  is  evident  from  the  Table~\ref{t1}  that the
contribution to the variance from the variation of mean TKE with fragment
mass  asymmetry,  $\sigma_E^2(kin)$,  increases  with  the  increase   in
excitatation  energy  and  mass  of the composite. For the lighter system
e.g. $O + Ag$, it is observed that this contribution is quite  large  due
to  the  asymmetric  nature  of  fission.  Further,  it  is seen from the
Table~\ref{t1}   that   there  is  a  very  good  agreement  between  the
$\sigma_E(th)$ and the  $\sigma_E$  extracted  from  experiment  for  the
systems  for  which the mass fragment yield data are available. Moreover,
the values $\sigma_E$, i.e., ( $\sigma_E^2(exp) - \sigma_E^2(kin)$ ), are
also shown in Fig.\ref{fig2} as open triangles and they agree very well  
with the predicted values of TKE variance.

\section{summary and conclusions}

We  have developed asymptotic expansion of the Fokker-Planck equation for
the systems in which the relative fluctuation of collective  variable  is
small  and  have  derived  the  equation  of  variances. The formalism is
applied to the case of fission where the  fluctuation  in  total  kinetic
energy  is small as compared to its mean value. The calculated variances,
$\sigma_E(th)$, underpredict the experimental data, $\sigma_E(exp)$.  The
discrepancy  between  the  two increases with the increase in the mass of
the composite and the excitation energy of  the  composite.  However,  we
emphasize  that  this  discrepancy  is  not  very  surprising because the
experimental values contain additional contributions  from  variation  of
the mean kinetic energy over full fragment mass yield distributions. Once
this  contribution is properly taken care of, the predicted TKE variances
are found to agree quite well with  the  $\sigma_E$  extracted  from  the
experimental  data for the systems where the fragment mass yield data are
available. Therefore, for a more direct test of theoretical models it  is
necessary  that  experimental  estimation  of  variances  should not have
admixture of other contributions arising due to  the  variation  of  mean
kinetic  energy  over  differnt  mass  yields.  This  may  be achieved if
measurements are done in smaller mass bins.

Thus,  it  may be concluded that the present approach is quite successful
in reproducing the extracted TKE variances  from  the  experimental  data
without  going in to solving the Fokker-Planck equation in detail. In the
present studies, the correlation of the  position  and  velocity  of  the
elongation  axis  has been found to be small. However, in the cases where
such condition is not valid the energy variance still can  be  calculated
by adding a term $2\mu\dot{r}(\frac{\partial(V_C+V_N)}{\partial r})c(t)$.

The  procedure  developed here could systematically generate higher order
hierarchies for relatively larger fluctuations than the ones  encountered
in  the  present studies. In those cases one may have to solve the higher
order  equation  which  would  involve  higher  order  derivatives of the
functions $h(x,y)$ and$H(x,y)$, in general.

\begin{figure}
\caption{Variation of $g(t)$, $c(t)$, $G(t)$ and $\sigma^2_E$ as a function
of time $t$ for the system $^{16}$O+$^{124}$Sn.}
\label{fig1}
\end{figure}

\begin{figure}
\caption{ Variation of $\sigma_E$ as a function of target mass number, for
 288 MeV $^{16}$O (upper half),  and 158.8 MeV $^{18}$O (lower half) induced
 fission reactions. Filled circles correspond to the experimental data
 \protect \cite{hinde} and solid lines correspond to the present theoretical
 results. Open triangles are the modified results using Eq. (3.3)
{\it (see text).}}
\label{fig2}
\end{figure}

\begin{figure}
\caption{ Variation of predicted values of $\sigma_E$ as a function of
fragment mass asymmetry, $a_{sym} = |A_1 - A_2|/ A_{CN} $.}
\label{sig_asym}
\end{figure}

\begin{table}
\caption{ Calculation of TKE variance}
\begin{tabular}{ccccccc}
System&E$_{lab}$ & $\sigma_E^2(exp)$  &
$\Sigma_{A_1}(<E(A_1,A_2)>-\bar{E})^2Y(A_1)$ \tablenote{See Eq.(3.3)}&
$\sigma_E^2$  & $\sigma_E$ & $\sigma_E(th)$  \\
&$(MeV)$ & $(MeV^2)$ & $(MeV^2)$ & $(MeV^2)$ & $(MeV)$ & $(MeV)$\\
\tableline
$^{18}$O + $^{154}$Sm & 158.8 & 112.3 & 10.1 & 102.3 & 10.1 & 9.7\\
$^{18}$O + $^{197}$Au & 158.8 & 190.4 & 15.2 & 175.2 & 13.2 & 11.9\\
$^{16}$O + $^{197}$Au & 288.0 & 331.2 & 49.7 & 281.5 & 16.7 & 13.3\\
$^{16}$O + $^{109}$Ag & 288.0 & 225.0 & 142.8 & 82.2 & 9.0 & 8.3\\
\end{tabular}
\label{t1}
\end{table}
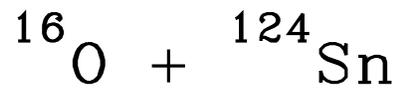
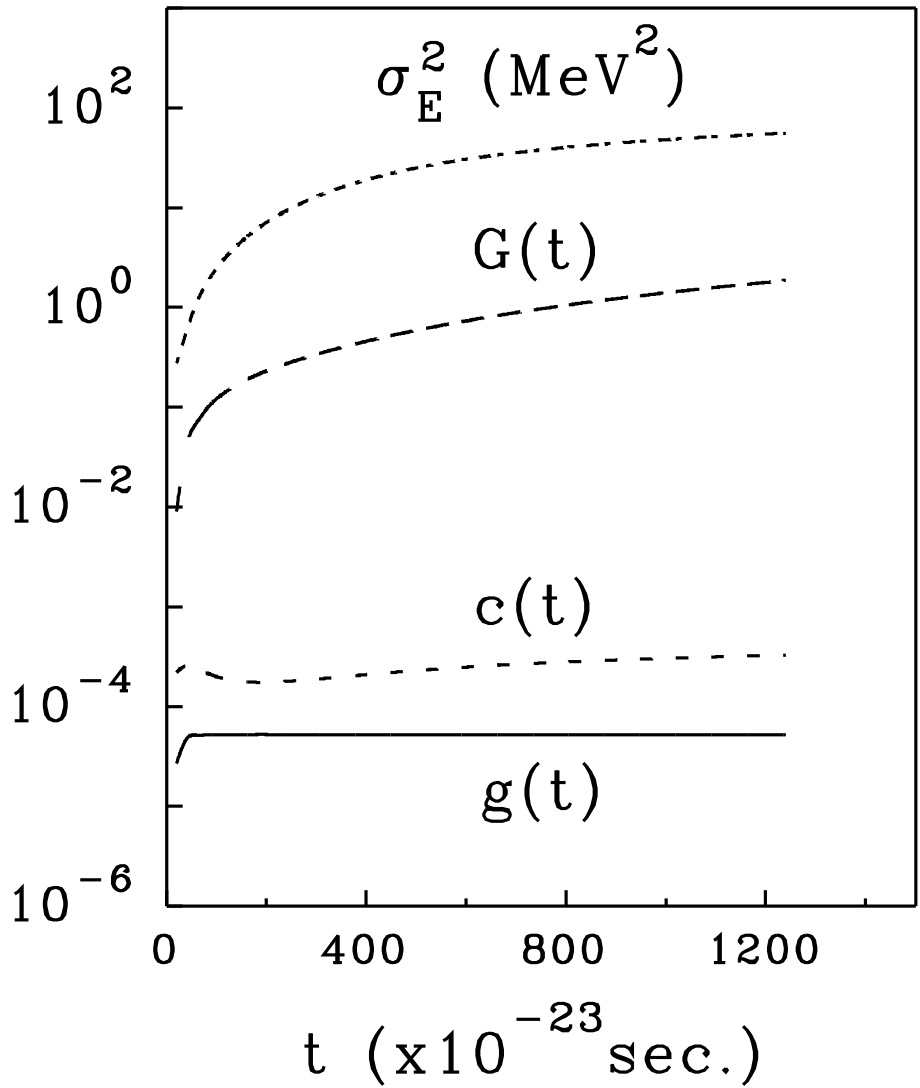
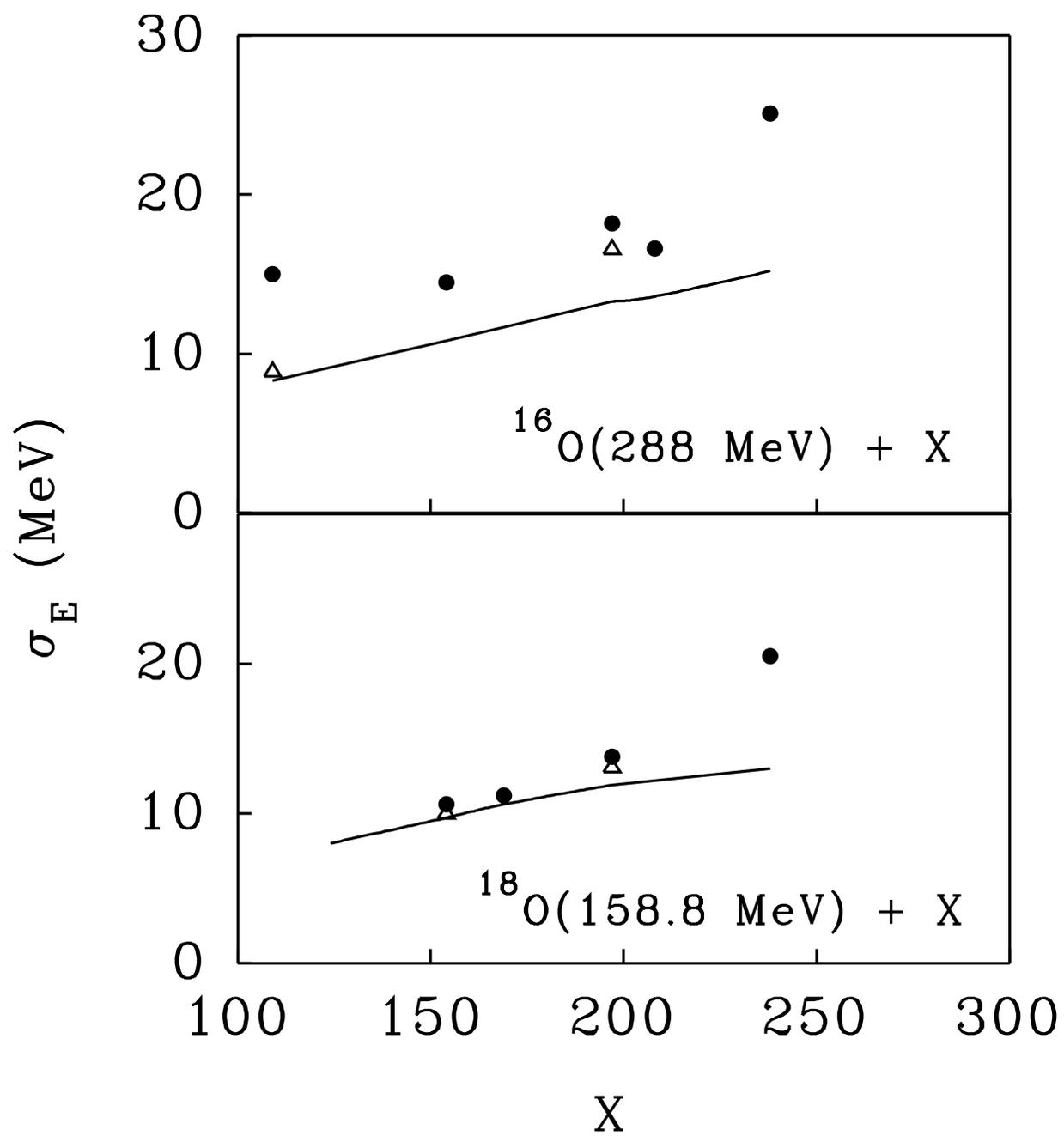
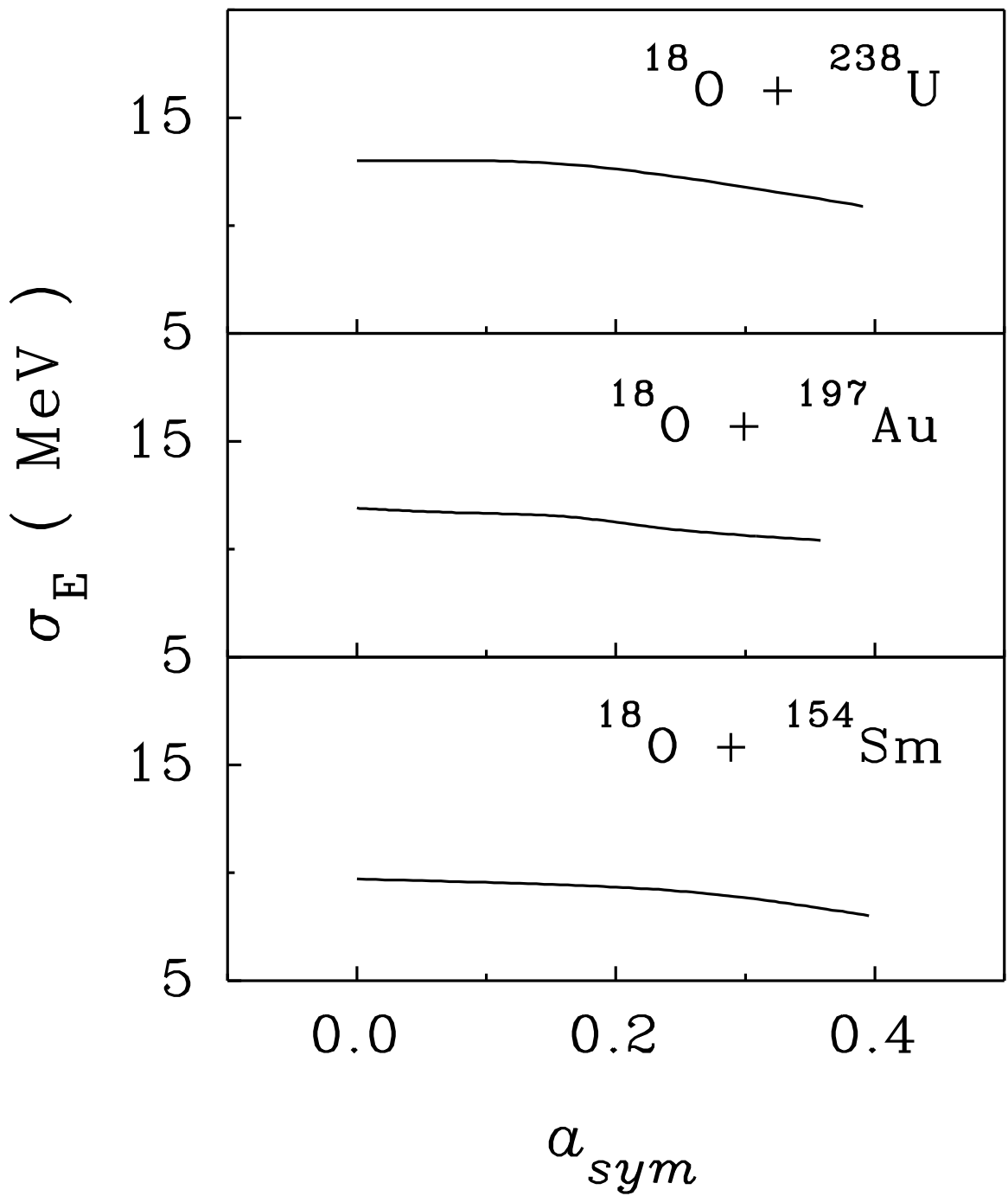
\end{document}